\documentstyle[prl,aps,multicol,epsfig,pstricks,amsmath,amssymb]{revtex}
\newcommand{\be}{\begin{equation}}
\newcommand{\ee}{\end{equation}}
\newcommand{\bq}{\begin{eqnarray}}
\newcommand{\eq}{\end{eqnarray}}

\def\qsqrt{{\sqrt{2} \kern-1.2em ^4}}
\def\CC{{\rm\kern.24em \vrule width.04em height1.46ex depth-.07ex
\kern-.30em C}}

\newcommand{\onecolm}{
  \end{multicols}
  \vspace{-3.5ex}
  \noindent\rule{0.5\textwidth}{0.1ex}\rule{0.1ex}{2ex}\hfill
}
\newcommand{\twocolm}{
  \hfill\raisebox{-1.9ex}{\rule{0.1ex}{2ex}}\rule{0.5\textwidth}{0.1ex}
  \vspace{-4ex}
  \begin{multicols}{2}
}

\def\PP{{\rm I\kern-.25em P}}
\def\RR{{\rm
         \vrule width.04em height1.58ex depth-.0ex
         \kern-.04em R}}
\def\id{{\rm 1\kern-.22em l}}
\def\ZZ{{\sf Z\kern-.44em Z}}
\def\NN{{\rm I\kern-.20em N}}

\begin{document}

\title{Integrable spin-boson interaction in the Tavis-Cummings model from a 
generic boundary twist}

\author{Luigi Amico$^{(a)}$ and Kazuhiro Hikami$^{(b)}$}

%\vspace*{0.2cm}

\address{$(a)$ Dipartimento di Metodologie Fisiche e Chimiche (DMFCI), 
        Universit\'a di Catania, viale A. Doria 6, I-95125 Catania, Italy\\
        MATIS $\&$ Istituto Nazionale per la Fisica della Materia, Unit\'a  di Catania, 
        Italy} 

\address{
$(b)$ Department of Physics, Graduate School of Science, University 
of Tokyo, Hongo 7-3-1, Bunkyo, Tokyo 113-0033, Japan}  

%\vspace*{-0.5cm}

%\pacs{02.30.Ik}{}
%\pacs{32.80.-t}{} 
%\pacs{03.75.Gg}{}
%

%\vspace*{-0.9cm}

\maketitle

\begin{abstract}
We construct models describing interaction between a spin $s$ and a single 
bosonic mode using a quantum inverse scattering procedure. The boundary conditions 
are generically twisted by generic matrices with both diagonal and off-diagonal entries.
The exact solution is obtained by  mapping the transfer matrix of the 
spin-boson system   
to  an auxiliary problem of a spin-$j$ coupled to the spin-$s$  with general twist of the 
boundary condition.  
The corresponding auxiliary transfer matrix  is diagonalized by a variation of   
the method of $Q$-matrices of Baxter. The exact solution of our problem is  obtained  applying certain large-$j$ limit to $su(2)_j$,  transforming it   into 
the bosonic algebra. 
 \end{abstract}

%%%%%%%%%%%%%%%%%%%%%%%%%%%%%%%%%%%%%%%%%%%%%%%%%%%%%%%%%%%%%%%%%%
%\begin{multicols}{2}

\vspace*{0.3cm}

{\it Introduction}\, 
Models representing interactions between  a single bosonic mode and spin degrees 
of freedom
find application in many different contexts. In atomic physics  they describe  
atoms interacting with  electromagnetic field~\cite{COHEN} and many phenomena like spontaneous 
emissions in cavity~\cite{KLEPPNER} and Rabi oscillations~\cite{REMPE} in two-level atoms 
are captured by an important 
representative of the models mentioned above, the Jaynes-Cummings (J-C) model~\cite{JAYNES}.   
Recently the J-C dynamics was intensively studied in the research field 
of ions in harmonic traps~\cite{ION-TRAPS} and then in quantum 
computation~\cite{RECENT-EXP}.
Finally this kind of models found   applications in quasi-2D 
semiconductors in transverse magnetic field~\cite{DATTA}. 
\\
Toy model Hamiltonians representing a single bosonic mode interacting with 
a spin $s$ are  of the following type
%\vspace*{-0.1cm}
\begin{equation}
H= H_{s-ph}+ 
\gamma \left (S^+ a^\dagger +S^- a \right ) 
\label{general-model}
\end{equation}
%\vspace*{-0.1cm}
where 
$H_{s-ph}=\omega a^\dagger a +  \alpha S^z + \beta 
\left (S^+ a +S^-  a^\dagger \right )$;
operators $a^\dagger$ and $a$ are bosonic operators that commute 
with spin operators $S^v$, $v=\{z,\pm\}$.  
The model~(\ref{general-model}) with $\gamma=\beta$ was 
originally proposed for quantum  optics purposes to describe a dipole-like  interaction
in  atoms-radiation systems and it is  known as  the Tavis-Cummings (T-C)
model\cite{TAVIS}; the model reduces 
to the J-C  one  for  $s=1/2$. 
In solid state physics models of type ~(\ref{general-model}) 
can describe certain quantum
circuits~\cite{BRANDES,FALCI}.

There are two cases in which the model can be solved analytically:
{\it i)}In the limit $s\rightarrow \infty$ there are
exact results~\cite{EMARY}. They were applied to study the  entanglement 
across the quantum phase 
transition between normal to super-radiant phase\cite{LAMBERT}.  
{\it ii) }For ``single photon'' interactions the model  can be simplified   
employing the  Rotating Wave Approximation (RWA) that neglects 
the so called ``counter rotating'' terms: $ S^+ a^\dagger$, $ S^-  a$.
Within the  RWA the model (\ref{general-model}) can be solved 
exactly\cite{JAYNES,TAVIS,LIEB}. 
For generic parameters $\alpha, \beta, \gamma, \omega$, and for finite 
$s$ the model, as it stands in Eq.~(\ref{general-model}), is non-integrable. 
Merging the model into the main stream of the   
Quantum Inverse Scattering (QIS) method\cite{KOREPIN-BOOK} constitues 
often a guide to discover unsuspected exactly solvable models 
with sufficiently generic interaction. 
According to this  method the Hamiltonian is obtained 
as output of the procedure that remarkably ensures the integrability  of the theory. 
In the simplest cases the QIS method provide integrable Hamiltonians   after 
periodic boundary condition are imposed. The variety of the integrable models 
can be considerably enriched by considering more general boundary 
conditions~\cite{SKLYANIN}. 
By this is meant that the 
monodromy matrix is multiplied by non-trivial 
matrices that ultimately cause the presence of boundary terms in the 
Hamiltonian. Of interest in the present paper is the case of constant 
boundary matrix; this realize the, so called, twisted  boundary conditions.
For the type of models under consideration the QIS method was employed 
in Ref.\cite{JURCO,BOGOLUBOV,RYBIN,KUNDU} where nonlinear generalizations of $H_{s-ph}$ 
were studied. These generalizations were obtained 
by twisting the boundary conditions. 
The twist matrices  were chosen  
as the same diagonal matrix for both the bosonic and 
spin degrees of freedom; finally in a certain sense (specified below)
they are classical. As a result, although nonlinear, these models 
contain the standard interaction. 
Here a more general interaction  is obtained 
applying more general boundary conditions: the twist matrices $K_B$ and $K_S$ 
are  non-diagonal, different for the boson and the spin, and 
of ``quantum'' nature (see  (\ref{quantum-boundary}))~\cite{DILORENZO}. 
The Hamiltonian we found is  Eq.~(\ref{our-model}). 
The  QIS method pave the way towards 
the exact solution of the theory through Bethe Ansatz.
In the cases where there 
exist an obvious ``reference'' state a direct (algebraic) Bethe Ansatz 
approach can be applied.
For the model we found here, however, there is no simple vacuum state since 
(\ref{our-model}) 
does not commute with $S^z+a^\dagger a $. A standard route to  attack the 
exact solutions  of such 
kind of spectral problems is to apply the technique that  
Baxter~\cite{BAXTER-BOOK} invented to obtain the  eigenvalues  
without  knowledge of the eigenstates. 
We obtain the eigenvalues (the calculation of the exact 
eigenstates will be the object of a future publication) in the following way.
We first define an auxiliary problem consisting of two spins 
with two distinct representations $j$ and $s$; 
the boundary conditions are {\it generically} twisted; 
the spin $j$ is affected by an ``impurity'' $\nu$. 
We diagonalize  the  auxiliary problem by adapting the Baxter method to it. 
Then the solution of the 
spin-bosonic problem is obtained performing certain $ j\rightarrow \infty $ limit 
(see (\ref{contraction})) in the results for the auxiliary 
spin-spin problem. 
 The eigenvalues  are given in~(\ref{eigenvalues}) and the parameters 
$\lambda_k$ are fixed by 
(\ref{be}). 

The paper is laid out as follows. 
In the next section we derive the integrable model. In the section III the exact 
solution is obtained. The section IV is devoted to our conclusions. 

{\it Integrability}.\,
The starting point of the QIS method is to define quantum Lax matrices  
$L(\lambda)$ and a scattering matrix $R(\lambda)$ satisfying the Yang Baxter (YB)
equation:
%\begin{equation}
$\displaystyle{
R(\lambda-\mu) L(\lambda)\otimes L(\mu)=L(\mu)\otimes L(\lambda) 
R(\lambda-\mu)\;,
}$
%\label{RL}
%\end{equation}
where $\lambda$ is the spectral parameter.
For the present case, the Lax operators $L$ we consider\cite{BOGOLUBOV} are 
\begin{equation}
L_S(\lambda):= \left(
\begin{array}{cc}
\lambda-\eta S^z &  \eta S^+   \\
\eta S^-  &\lambda+\eta S^z   
\end{array}
\right) \; ,
\end{equation}
\begin{equation}
L_B(\lambda):= \left(
\begin{array}{cc}
\lambda -\Delta - \eta^{-1}- \eta a^\dagger a  &  a^\dagger   \\
a  & - \eta^{-1} 
\end{array}
\right) \; ,
\label{bosonic-Lax}
\end{equation}
each satisfying the YB equation with: 
$
%\begin{equation}
R(\lambda;\eta)=\eta \id \otimes \id + \lambda \PP  \;, 
%\label{R-matrix}
%\end{equation}
$
where $\eta \in \RR$ and  $\PP$ is the permutation:  
$\PP A\otimes B \PP =B\otimes A $. 
The monodromy matrix is 
\begin{equation}
T(\lambda)=K_B L_B(\lambda) K_S L_S(\lambda)\;,
\label{monodromy}
\end{equation} 
where $K_B$ and $K_S$ are $\CC$-number matrices 
that  produce boundary terms (without ``internal  dynamics'',  
the  matrices $K$ not depending  on $\lambda$). 
Notice that we have two different 
boundaries  each for the spin and for the boson. 
In Refs.\cite{BOGOLUBOV,RYBIN}
$K_B\equiv K_S$ is assumed;
the T-C model (without counter rotating terms) $H_{s-ph}$ 
is obtained for $\eta\rightarrow 0$.
The matrix $T(\lambda)$ fulfills the YB 
relation:  
$
%\begin{equation}
R(\lambda-\mu) T(\lambda)\otimes T(\mu)=T(\mu)\otimes T(\lambda) 
R(\lambda-\mu)\;,
$
%\end{equation}
due to the fact that $[R,K_B\otimes K_B]=[R,K_S\otimes K_S]=0$ holds
for any numeric matrix because of  the $sl(2)$ symmetry of the $R$-matrix.
The transfer matrix is defined as
$
t (\lambda):=tr_{(0)} T(\lambda) \;  
$
where $tr_{(0)}$ means trace  in the auxiliary 
space. $t(\lambda)$ is a generating functional of integrals of motion since:
%it commutes with itself  at different values of spectral parameters:
$[t (\lambda),t (\mu)]=0$. 
For the present case the transfer matrix 
can be chosen as a polynomial in $\eta$: $t(\lambda) = \sum_{l=-g}^{h} \eta^l 
t_l(\lambda)$; then 
the coefficients of $ [t(\lambda),t(\mu)]=\sum_{l=-2 g}^{2 h} \eta^l 
C_l(\lambda,\mu)$ vanish at any $\eta$-power. 
The following assumption is crucial for our purposes:
The entries of the matrices $K$ depend on  
$\eta$  
\begin{equation}
\label{quantum-boundary}
K_{X\,ij}=K_{X\,ij}^{(0)}+K_{X\,ij}^{(1)} \eta + \dots\;, \qquad X=\{B,S\}.
\end{equation}
The parameter $\eta$   is usually 
called ``quantum parameter'' 
since it controls the limit how to recover the classical scattering matrix $r(\lambda)$ from 
the matrix $R(\lambda)$. In this sense our  matrices $K$ in 
(\ref{quantum-boundary}) describe ``quantum 
systems'' which we couple to the the boson and to the spin at the boundary 
(twist matrices 
that are independent on $\eta$ might be considered as classical boundaries). 
In brief, the main idea of our procedure is to play with the boundaries 
$K$ in such a way that 
$C_l(\lambda,\mu)=0 \, \Leftrightarrow \, [t_l(\lambda),t_l(\mu)]=0$ for certain $l$. 
In order to obtain the model we are interested in, 
the degree  and the coefficients of the polynomials are fixed such that:
{\it i)} $t_m(\lambda)$ describes an integrable model, then $t_l(\lambda) $ must be  $\CC$-numbers for all $l<m$; 
{\it ii)} the model results  containing the  
counter-rotating terms; {\it iii)} the obtained operator is Hermitian.     
All these conditions translate in a system of equations for  the 
entries of the matrices $K$; we shall see that these parameters 
will be  the coupling constants of the Hamiltonian.
It turns out to be sufficient to consider the entries of the $K$ matrices 
to be  linear 
in $\eta$. Such entries are restricted  to 
\begin{eqnarray}
K_{B\,11}^{(0)}= K_{B\,22}^{(0)} \; , \; K_{S\,11}^{(0)}= K_{S\,22}^{(0)} \; ,
 K_{B\,21}^{(0)}= K_{B\,12}^{(0)} \; , \; {K_{S\,12}^{(0)}}={K_{S\,21}^{(0)}} \;,
K_{S\,12}^{(0)}= -\frac{K_{B\,21}^{(0)}K_{S\,22}^{(0)}}{K_{B\,11}^{(0)}}  \;,
\end{eqnarray}
\begin{eqnarray}
K_{S\,11}^{(1)}=\frac{K_{B\,22}^{(1)}K_{S\,22}^{(0)}}{K_{B\,11}^{(0)}} +
\frac{K_{B\,11}^{(0)}}{K_{B\,12}^{(0)}} (K_{S\,21}^{(1)}-K_{S\,12}^{(1)})- 
\frac{K_{S\,11}^{(0)}}{K_{B\,12}^{(0)}} (K_{B\,12}^{(1)}-K_{B\,21}^{(1)})+
\frac{K_{S\,22}^{(0)}}{K_{B\,11}^{(0)}} K_{B\,11}^{(1)}-K_{S\,22}^{(1)} \;.\nonumber 
\end{eqnarray}
We take $t_1(\lambda)$ as    the Hamiltonian
%\onecolm
%\vspace*{-0.7cm}
%\begin{widetext}
%\vspace*{-0.5cm}
%\begin{multline}
\begin{eqnarray}
t_1(\lambda) \doteq  H
  =
  W(\lambda) S^z +
  \lambda(U+V) a^\dagger a  
  + 2 \left [Y + \sqrt{U V} (\Delta- \lambda) \right ] S^x +Z \lambda(a+a^\dagger)
  \\
  -U(a S^+ + a^\dagger S^-) +
  V(a^\dagger S^+ + a S^-)  - 2 \sqrt{U V} ( a+ a^\dagger)  S^z \;,
\label{our-model}
%\end{multline}
%\end{widetext}
%\twocolm
%\vspace*{-1cm}
\end{eqnarray}
where the couplings are 
\begin{eqnarray}
\label{parameters}
\hspace*{-1cm} &&  \hspace*{-1cm}
W(\lambda)= -\frac{(U+V) Z}{\sqrt{UV}} + (V-U) (\Delta - \lambda-\frac{Y}{\sqrt{UV}}) \;,  \\
&&\hspace*{-1cm}U=-K_{B\,22}^{(0)}K_{S\,22}^{(0)}\; , \; V=-{K_{B\,22}^{(0)}(K_{S\,21}^{(0)})^2}/{K_{S\,22}^{(0)}} \\
&&\hspace*{-1cm}Y=K_{B\,21}^{(0)}   \left ({K_{B\,11}^{(1)}K_{S\,21}^{(0)}}/{K_{B\,22}^{(0)}}-K_{S\,21}^{(1)}\right )-K_{B\,12}^{(0)} K_{S\,22}^{(1)}-K_{B\,22}^{(0)} K_{S\,12}^{(1)}  \;, \\
&& \hspace*{-1cm}Z=Y+ K_{B\,22}^{(0)} \left (K_{S\,12}^{(1)}+K_{S\,21}^{(1)}\right ) + 2 K_{B\,22}^{(1)} K_{S\,21}^{(0)} \;.
\end{eqnarray}
The 
coupling constants obey~(\ref{parameters})~for the model to be integrable;
parameters $\Delta$,  $\lambda$, $X$, $Z$  can be set freely; 
the quantity $UV$ 
must be positive. 
Nevertheless the rotating and counter-rotating terms can be
adjusted to have the same 
sign by  acting 
on the operators:
$S^-\rightarrow S^- e^{i\frac{\pi}{2}}$ and 
$a\rightarrow a e^{i\frac{\pi}{2}}$; 
the third, fourth and last terms in Eq.~(\ref{our-model}) 
are transformed accordingly.        
We observe that the simultaneous presence of rotating and counter-rotating 
terms  preserves the integrability only if 
a further term $(a+a^\dagger ) S^z$ appears 
in the model (the term $a+a^\dagger$ can be transformed out  by a 
translation: $a\rightarrow a+\xi$ with $\xi=-Z/(U+V)$; the coefficients of 
$S^x$ and $S^z$ are  shifted by $\xi (U-V)$ and $-2 \xi \sqrt{UV}$ 
respectively). 
Restricting the boundary conditions: $K_B=K_S$ (non-diagonal)
induces the further constraint $U=V$. 
In this case our model reduces  to the  
Rashba Hamiltonian 
in a constant magnetic field~\cite{RASHBA}
(the bosonic number labeling  the Landau levels; see also Ref.\cite{CLIVE}).

The constants of  motion of (\ref{our-model}) are 
$t_1(0)$ and $\partial_\lambda t_1(\lambda)$ (only two of $H$, 
$t_1(0)$, $\partial_\lambda t_1(\lambda)$ are independent).~$\partial_\lambda t_1(\lambda)$ can be easily diagonalized
: ${\cal U} \partial_\lambda t_1(\lambda) {\cal U}^{-1}|\phi\rangle=
 ( (U+V)(n+m)  -M_- ) |\phi \rangle$. In the $|\phi\rangle $ basis   the 
Hilbert space of the Hamiltonian blocks into invariant  subspaces
labelled by the  
bosonic number $n \geq 0$ and with  $S^z|\phi\rangle =m |\phi\rangle$;
$M_-\doteq Z^2 /(U+V)$.
This will be used  to classify the excitations in the Bethe equations 
(\ref{be}).

{\it Exact eigenvalues}.
To diagonalize the model (\ref{our-model}) we define an auxiliary 
inhomogeneous spin problem. 
We use the property that  a spin $j$-$su(2)$  can be contracted to the 
Weyl-Heisenberg algebra~\cite{GILMORE} through the singular limit $\varepsilon\rightarrow \infty $ 
of a Dyson-Maleev  transformation   
\begin{equation}
\left \{-\eta J^-, \; \frac{1}{\eta \varepsilon^2} J^+, \; J^z\right \}
\mapsto \left \{a^\dagger, \; a, \; -a^\dagger a -\frac{ \varepsilon^2}{2} \right \}\; ;
\label{contraction}
\end{equation}
such a limit corresponds to $j=-\varepsilon^2/2 \rightarrow \infty$. 
The bosonic Lax matrix is thus expressed as limit of a spin-$j$ Lax matrix:
\begin{equation}
{L}_{B}(\lambda) =  \lim_{\varepsilon \rightarrow  \infty}
 (-\frac{1}{\eta \varepsilon}) K(\varepsilon) \sigma_y\sigma_z {L}_{J}(\lambda-\nu)\sigma_y\; ,
\end{equation}
where $K(\varepsilon)=diag\left \{\eta \varepsilon,(\eta \varepsilon)^{-1}\right \}$
and 
\begin{equation}
{L}_{J}(\lambda-\nu) =\left(
\begin{array}{cc}
\lambda -\nu-\eta J^z &  \eta J^+   \\
\eta J^-  &\lambda-\nu +\eta J^z   
\end{array}
\right) \;,
\end{equation}
the ``inhomogeneity'' parameter being    set to 
\begin{equation}
\nu=-\eta \varepsilon^2/2 +\eta^{-1}+\Delta \;.
\label{imp}
\end{equation}
Thus the monodromy matrix Eq~(\ref{monodromy}) can be written as 
$
T(\lambda)=\lim_{\epsilon\rightarrow \infty } T_a(\lambda)
$
where $T_a$ is an auxiliary monodromy matrix defined as
\begin{equation}
T_a\doteq K_J L_J(\lambda-\nu) \sigma_y K_S L_S(\lambda)\;,
\label{monodromy-auxiliary}
\end{equation}
with 
$
K_J\doteq 
-1/(\eta \varepsilon)  K_B K(\epsilon) \sigma_y\sigma_z 
$.
$t_a=tr_0\{T_a\}$ can be diagonalized adapting  
the Baxter method~\cite{BAXTER-BOOK} for off-diagonal twisted spin-$s$ 
chain~\cite{YUNG}. In the 
present case the ``chain'' consists of only two sites; the twist matrices are distinct and containing 
both diagonal and off-diagonal entries.  The details of the calculations will be reported 
elsewhere.
%%%%%%%%%%%%%%%%
 The  Baxter equation reads
%\begin{widetext}
%\onecolm
\begin{eqnarray}
t_a(\lambda)  Q(\lambda) = R^{(\mp)}\; 
  (\lambda -s \eta) (\lambda-\nu-j \eta ) Q(\lambda+\eta )
%-\left ({\det{K_J} \det{K_S}}/
+{R^{(\pm)}}
\,  (\lambda+s \eta) (\lambda-\nu+j \eta) 
 Q(\lambda-\eta)\;, 
\label{baxter-auxiliary}
\end{eqnarray}
where $R^{(\pm)}\doteq \frac{1}{2}\left [tr (K_J \sigma_y K_S) \pm \sqrt{(tr 
(K_J \sigma_y K_S))^2+4 \det (K_J K_S)} \right ]$. The quantities  
$Q(\lambda)$ are $(2s+2j+2)\times (2s+2j+2)$ matrices and constructed in 
the standard way\cite{BAXTER-BOOK,YUNG}; they fulfill 
$[Q(\lambda),Q(\mu)]=[Q(\lambda),t_a(\lambda)]=0$. Eq~(\ref{baxter-auxiliary}) 
fixes the eigenvalue $\tau_a(\lambda)^{(\pm)}$
of the transfer matrix 
\begin{eqnarray}
  \tau_a(\lambda)^{(\pm)}=R^{(\mp)}\; 
  (\lambda -s \eta) (\lambda-\nu-j \eta ) 
 && \prod_{i=1}^{2s+2j} 
  \frac{\lambda-\lambda_i+\eta}
  {\lambda-\lambda_i} \nonumber \\
 &&+ 
  R^{(\pm)}  \, (\lambda+s   \eta)
  (\lambda-\nu+j \eta) 
  \prod_{i=1}^{2s+2j}
  \frac{\lambda-\lambda_i-\eta}
  {\lambda-\lambda_i} \;,
\end{eqnarray}
where the variables $\lambda_j$ are solutions of the  equations
\begin{equation}
 \frac{  R^{(\mp)} }{R^{(\pm)}}
  \frac{(\lambda_k -s \eta) (\lambda_k-\nu-j \eta )}
  {(\lambda_k+s\eta) (\lambda_k-\nu+j\eta)}
  =
  \prod_{i =1 \atop i\neq k}^{2s+2j}
  \frac{\lambda_k-\lambda_i-\eta}
  {\lambda_k-\lambda_i+\eta} \; ,\qquad \qquad\; k=1,\dots, 2(s+j) \;.  
  \label{auxiliary-be}
\end{equation}
To obtain the solution of the bosonic problem 
we perform the algebraic 
contraction at the level of Eqs.~(\ref{auxiliary-be}). 
This can be done expanding ${R^{(\pm)}=\sum_{m} 
R^{(\pm)}_{2m}/(\varepsilon \eta)^{2m}}$ and 
taking into account of Eq.~(\ref{imp}); a generalizations of the Bethe equations found in~\cite{BOGOLUBOV} are obtained.
By a further expansion $R^{(\pm)}_{2m}=\sum_{n=0}^2 R^{(\pm)}_{2m,n}\eta^{n}$ in the latter equations, 
the eigenvalues ${\cal E}$ of the model 
(\ref{our-model})  can be obtained as their linear terms
in $\eta$: 
\begin{eqnarray}
{\cal E}={\cal E}_0(\lambda)+\sum_{j=1}^{2s+n} \frac{1}{\lambda-\lambda_j}\left[- \lambda 
x^{-}_0\sum_{l\neq j}^{2s+n} \frac{1}{\lambda_j-\lambda_l}-\lambda(\lambda-\Delta) y^{-}_0-\lambda x^{+}_1 +s x^{-}_0\right ] \;,
\label{eigenvalues} 
\end{eqnarray}
%\end{widetext}
%\twocolm 
where ${\cal E}_0(\lambda)=( \lambda-\Delta)(\lambda y^{+}_1
+ s y^{-}_0) -\lambda x^{-}_2 +s x^{+}_1$; with 
$x^{\alpha}_n=R^{(\mp)}_{0,n}   -\alpha  R^{(\pm)}_{0,n}- 
R^{(\mp)}_{2,n}$ and $y^{\alpha}_n=R^{(\pm)}_{0,n}+ \alpha R^{(\mp)}_{0,n}$ $\alpha=\{+,-\}$. 
The quantities $\lambda_k$ are fixed by   
\begin{equation}
\frac{s x^{-}_0}{\lambda_k}-\left (\lambda_k -\Delta\right ) y^-_0=x^-_0\sum_{l\neq k}^{2s+n} \frac{1}{\lambda_k-\lambda_l}+x^{+}_1\, ,  
\label{be}
\end{equation}
where $ k=1,\dots, 2s +n$;
the quantum number 
$n$ labels the excitations as discussed before. For generic $2s+n$ the equations above 
can be solved numerically. Alternatively the quantities $\lambda_k$ can 
be obtained as roots of the polynomial $P(\lambda)=\prod_{s=1}^{2s+n} 
(\lambda-\lambda_s)$ satisfying 
\begin{equation}
\lambda P''(\lambda)+\frac{2}{x^-_0} 
\left [y^-_0 \lambda^2 +( y^-_0 \Delta+x^+_1)\lambda- sx^-_0\right ]P'(\lambda)-
\left [\zeta -(2s +n)\lambda\right ]P(\lambda)=0 
\end{equation}
where $\zeta$ is fixed by imposing that $\lambda=0$ is a simple root of 
$\lambda P(\lambda)$: $\zeta=s x^-_0\frac{P'(0)}{P{0}}$~\cite{BOGOLUBOV}. 

\begin{figure}
\vspace*{-0.3cm}
\begin{center}
\includegraphics[width=8.5cm]{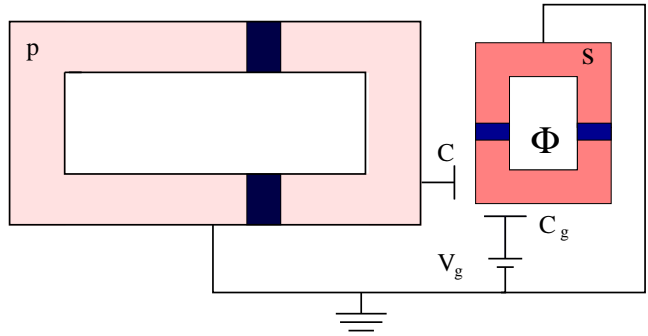}\\
\caption{The quantum circuit described by the Hamiltonian (\ref{H-circuit}). 
The primary device $p$ is a resonant circuit controlling  the flux-qubit
$s$ by the inductive coupling caused by  $\Phi=\phi_{ext}+L_p I_p$.}
\label{twosquid}
\end{center}
\end{figure}

{\it Conclusions}.  
By the Quantum Inverse scattering method we have  constructed integrable 
T-C models with twisted boundary conditions.
The twist matrices are generic in the sense that they contain both diagonal and 
non-diagonal entries. They are responsible for the presence of     
rotating and counter-rotating terms in the  Hamiltonian.
The spectrum  is  computed through the Baxter method. As far as we know this method 
is applied to spin-boson systems for the first time; the subtleties related to 
the bosonic limit, recovered for infinite spin length are dug out.   
Integrability  and exact solution can be obtained provided that a 
further term $ \propto (a+a^\dagger) S^z$  is considered. Interestingly enough 
we found a global 
rotation of the spin/bosonic degrees of freedom such  that the rotating 
terms (alternatively, the counter-rotating 
terms) are compensated  out\cite{MARTINS}.   
We conjecture that ``true'' counter-rotating terms in the Tavis-Cummings 
model could be inserted considering dynamical boundaries: 
$K_X=K_X(\lambda)$; alternatively one should consider $XYZ$ symmetry of the 
scattering matrix.
These terms  serve to a reliable description of certain 
systems in quantum optics\cite{NON-CLASSICAL-EXP,NON-CLASSICAL-THEORY} or to model the spin-orbit 
interaction  in heterostructures where the simultaneous Rashba and 
Dresselhaus terms are important\cite{SPINTRONICS}.  Our paper 
could pave the way to costruct integrable Hamiltonians for such  
physical situations. 
As immediate application, we notice that the Hamiltonian (\ref{our-model}) 
describes the quantum circuit of Fig.(\ref{twosquid}).  
Two coupled 
dc-Superconducting Quantum Interference Devices (SQUIDs) are  
coupled inductively. %with mutual inductance $M$. 
The primary device $p$ is intended built  with large Josephson junctions to be described by 
a classical SQUID Hamiltonian\cite{TINKHAM} (whose degrees of freedom are, then,  bosonic) 
flowed by the current $I_p$ (this circuit plays the role of the $LC$ resonant circuit of 
Ref.~\cite{FALCI}); 
the secondary SQUID $s$, with small junctions, is accommodated inside the 
primary
and pierced by the magnetic flux: $\Phi=\phi_{ext}+L_p I_p$
($L_p$ is the inductance of the circuit). Thus the   
secondary is a quantum SQUID controlled by the 
classical one. 
The effective Josephson coupling of the quantum SQUID 
depends on the flux $E_J^s (\Phi)\simeq E_J^s (\phi_{ext})+\tilde{L}_p I_p$ . 
This kind of setups are intensively studied as controllable 
flux-qubits\cite{CHIORESCU,MURALI} to data-bus transferring in many protocols of  
quantum computation\cite{FALCI}. 
The circuit  Hamiltonian is
\begin{equation}
{\cal H}_{circuit}=\omega_p a^\dagger a -
2 E^s_{J} (\phi_{ext})  S^x  
%+2 (E^s_C-E_g) S^y 
-2 \tilde{L}_p(a+a^\dagger) S^x
- i M (a-a^\dagger) S^y +2 V_C  (a+a^\dagger) S^z
\label{H-circuit}
\end{equation}
where $\omega_p$  is the ``frequency'' of the  primary SQUID (or the natural frequency 
of the resonant circuit\cite{FALCI}), $V_C$ is due to the capacitive 
coupling between the SQUID's; $E_J^\alpha,\;E_C^\alpha\;, \alpha=\{p,s\}$ are related respectively 
to the Josephson and the charging energies of the junctions and $M$ is the mutual inductance;
the gate voltage  $V_g$  is tuned to the charge degeneracy point\cite{TINKHAM}. 
For generic  circuit-parameters 
the dynamics of the qubit is intricated by the presence of the counter-rotating 
terms, making the device not reliable in the communication protocols. 
Our calculation suggests how the circuit parameters can be tuned to reproduce 
our model (\ref{our-model}); for it the dynamics is not altered by the 
presence of the counter-rotating terms. Using this trick  the qubit  dynamics 
can be  effectively ``protected''  at {\it any frequency $\omega_p$}.     
The relation between the 
circuit-parameters and coefficients in the Hamiltonian (\ref{our-model}) is:
$\{\omega_p,E^s_J, 2 \tilde{L}_p, M, V_C \} \rightarrow 
\{\lambda (U+V), 2 Y , U+V, U-V , -2 \sqrt{UV}\}$ ($Z=0$ is set for simplicity),
implying that $V_C$ 
should be tuned to $V_C=\sqrt{M^2-4 \tilde{L_p}^2}/2$ to make 
un-effective the counter-rotating terms. 
\vspace*{-0.3cm}
\acknowledgements
\vspace*{-0.2cm}
Discussions with G. Falci  and A. Osterloh are acknowledged. We are grateful to D. Averin, 
P. Kulish, E. Paladino, R. Fazio and R. Richardson   
for helpful comments. While completing this paper, equations similar 
to~(\ref{baxter-auxiliary})-(\ref{auxiliary-be}) appeared  in \cite{RIBEIRO}.

%\end{multicols}
\end{document}